\documentclass[sigconf]{acmart}
\usepackage{booktabs} % For formal tables
\usepackage[normalem]{ulem}
\usepackage{multirow}
\usepackage{bm}
\useunder{\uline}{\ul}{}
\usepackage{subcaption}
\usepackage[para]{footmisc}
\usepackage{enumitem}
\usepackage{balance} % to balance the references

\hypersetup{
,bookmarksnumbered=true%
,hypertexnames=false%
,breaklinks=true%
,colorlinks=true%
,linkcolor=black%
,citecolor=black%
,urlcolor=black%
}

\definecolor{tealblue}{rgb}{0.21, 0.46, 0.53}
\definecolor{wildstrawberry}{rgb}{1.0, 0.26, 0.64}
\definecolor{ao(english)}{rgb}{0.0, 0.5, 0.0}

\setcopyright{acmcopyright}
\copyrightyear{2019}
\acmYear{2019}
\setcopyright{acmcopyright}
\acmConference[CHIIR '19]{2019 Conference on Human Information Interaction and Retrieval}{March 10--14, 2019}{Glasgow, Scotland Uk}
\acmBooktitle{2019 Conference on Human Information Interaction and Retrieval (CHIIR '19), March 10--14, 2019, Glasgow, Scotland Uk}
\acmPrice{15.00}
\acmDOI{10.1145/3295750.3298924}
\acmISBN{978-1-4503-6025-8/19/03}

\def\ignore#1{}

\newcommand{\lycomment}[1]{\textcolor{red}{[#1---ly]}}

\newcommand{\jtcomment}[1]{\textcolor{orange}{[#1---jt]}}

\begin{document}
\title{User Intent Prediction in Information-seeking Conversations}

\begin{abstract}

    Conversational assistants are being progressively adopted by the general population. However, they are not capable of handling complicated information-seeking tasks that involve multiple turns of information exchange. Due to the limited communication bandwidth in conversational search, it is important for conversational assistants to accurately detect and predict user intent in information-seeking conversations. In this paper, we investigate two aspects of user intent prediction in an information-seeking setting. First, we extract features based on the content, structural, and sentiment characteristics of a given utterance, and use classic machine learning methods to perform user intent prediction. We then conduct an in-depth feature importance analysis to identify key features in this prediction task. We find that structural features contribute most to the prediction performance. Given this finding, we construct neural classifiers to incorporate context information and achieve better performance without feature engineering. Our findings can provide insights into the important factors and effective methods of user intent prediction in information-seeking conversations. 

\end{abstract}

% <concept_id>10002951.10003317</concept_id>
% <concept_desc>Information systems~Information retrieval</concept_desc>
% <concept_id>10002951.10003317.10003347.10003348</concept_id>
% <concept_desc>Information systems~Question answering</concept_desc>
% <concept_id>10010147.10010178.10010179.10010181</concept_id>
% <concept_desc>Computing methodologies~Discourse, dialogue and pragmatics</concept_desc>
% <concept_id>10010147.10010257</concept_id>
% <concept_desc>Computing methodologies~Machine learning</concept_desc>
% <concept_id>10010147.10010257.10010293.10010294</concept_id>
% <concept_desc>Computing methodologies~Neural networks</concept_desc>

% We no longer use \terms command

\keywords{User Intent Prediction; Information-seeking Conversations; Conversational Search; Multi-turn Question Answering}

\author{Chen Qu}
\affiliation{
	\institution{University of Massachusetts Amherst}
}
\email{chenqu@cs.umass.edu}

\author{Liu Yang}
\affiliation{
	\institution{University of Massachusetts Amherst}
}
\email{lyang@cs.umass.edu}

\author{W. Bruce Croft}
\affiliation{
	\institution{University of Massachusetts Amherst}
}
\email{croft@cs.umass.edu}

\author{Yongfeng Zhang}
\affiliation{
	\institution{Rutgers University}
}
\email{yongfeng.zhang@rutgers.edu}

\author{Johanne R. Trippas}
\affiliation{
	\institution{RMIT University}
}
\email{johanne.trippas@rmit.edu.au}

\author{Minghui Qiu}
\affiliation{
	\institution{Alibaba Group}
}
\email{minghui.qmh@alibaba-inc.com}
%\authornote{}
%\orcid{}
% \email{authors@authors.com}

\affiliation{%
  \institution{}
  \streetaddress{}
  \city{}
  \state{}
  \postcode{}
}

\maketitle

%----------------------------------------------------------------------------------------
\section{Introduction}
\label{sec:intro}
%---------------------------------------------------------------------------------------

% Paragraph 1-2: What is the task ?
% [P1] Start with broader background

Many companies have launched conversational assistants (CA) such as Amazon Echo, Google Home, Microsoft Cortana, etc. These devices allow users to issue voice commands to the CA for goal oriented tasks or to conduct simple question answering (QA) tasks, such as adding calendar events or asking for news.
%Such conversation oriented interfaces become increasingly popular among users. 
This trend has led many researchers in the information retrieval (IR) and natural language processing (NLP) community to pay more attention to conversational search. For examples, SIGIR'18, ICTIR'17, and EMNLP'18 all have workshops\footnote{\url{https://sites.google.com/view/cair-ws/cair-2018}}\textsuperscript{,}\footnote{\url{http://sigir.org/ictir2017/sessions/search-oriented-conversational-ai-scai/}}\textsuperscript{,}\footnote{\url{https://scai.info/}} on conversational search.

\if0
However, most CAs are not yet capable of handling multi-turn information-seeking conversations, where users have multiple rounds of information exchange with CAs to retrieve answers. This kind of conversations can be best described as a joint effort between human users and CAs in an exploratory manner. Different from single-turn QA, conversational QA requires CAs to proactively ask or confirm before presenting an answer~\cite{DBLP:journals/corr/LiMCRW16a} and expect user feedback after presenting an answer. Despite the effort of all the major internet companies, iterative answer finding with CAs has not yet been achieved. A main reason is the failure to model the conversation bout the information need both before and after an answer has been given. 
The key to the solution is to accurately detect and predict user intent in this interactive information-seeking process. Concretely, CAs would be capable to improve previous answers if it can correctly process critical information provided by the users, such as relevance judgment (feedback) and clarifications of the information need. Moreover, we expect CAs to request more information proactively when they are not confident to provide an answer. 
% \lycomment{It is better to use some related works to support your claim. For example, we can cite this paper \cite{DBLP:journals/corr/LiMCRW16a}  to show that we can train a CA which can proactively ask questions if she is not confident on the answer.}
\fi

However, most CAs are not yet capable of handling multi-turn information-seeking conversations, where users have multiple rounds of information exchange with CAs to retrieve or specify answers. 
%This kind of conversations can be best described as a joint effort between human users and CAs in an exploratory manner. 
%Different from single-turn QA, conversational QA requires CAs to proactively ask or confirm before presenting an answer~\cite{DBLP:journals/corr/LiMCRW16a} and expect user feedback after presenting an answer. 
%Despite the effort of all the major internet companies, iterative answer finding with CAs has not yet been achieved. 
One reason is the difficulty to model the conversation about the information need both before and after an answer has been given. Thus an important step in modeling conversational interactions is to accurately detect and predict user intent in this interactive information-seeking process. More specifically, CAs should be capable of improving previous answers if they can correctly process critical information provided by the users, such as relevance judgments (feedback) and clarifications of the information need. Thus, CAs need to elicit more information proactively when they are not confident, before providing an answer~\cite{Trippas:2018:IDS:3176349.3176387}.

The Learn-IR workshop\footnote{\url{https://task-ir.github.io/wsdm2018-learnIR-workshop/}} at WSDM'18 highlighted the significant research need for user intent analysis and prediction in an interactive information-seeking process. To address this research demand, we conducted experiments on user intent prediction using the MSDialog~\cite{Qu2018UserIntent} data. This data collection consists of multi-turn information-seeking dialogs in the technical support domain. The dialogs are typically initiated by an information seeker who asks technical issues about Microsoft products, such as ``How do I downgrade from Windows 10 to Windows 7?''. This kind of question is non-factoid and often requires multiple rounds of conversational interactions. The answers are provided by Microsoft staff and other experienced product users such as ``Microsoft Most Valuable Professionals'' (MVPs).
% \footnote{\url{https://mvp.microsoft.com/en-us/Overview}}. 
These human agents (information providers) also explicitly ask for feedback on their provided answers and thus keep the users engaged. The MSDialog data is annotated with a set of 12 \textit{user intent types}~\cite{DBLP:conf/webdb/BhatiaM12, Qu2018UserIntent}. There are different definitions of ``user intent'' in our field. In this paper, user intent refers to a taxonomy of utterances in information-seeking conversations (Table~\ref{tab:taxonomy}). Each utterance of the MSDialog was annotated with the user intent types using Amazon Mechanical Turk (MTurk)\footnote{\url{https://www.mturk.com/}}. MSDialog provides a high-quality resource to show how information-seeking conversations are structured between humans. In addition to MSDialog, we also used a portion of Ubuntu Dialog Corpus (UDC)~\cite{Lowe2015The} which was annotated with the same user intent types in~\cite{Qu2018UserIntent} to further validate our findings.
%CAs have the potential to outperform human agents in the future, but at the time being we expect CAs to learn and imitate the behaviors of human agents.
% \jtcomment{Could you explain the last sentence? } \lycomment{I suggest to remove the last sentence. The performance current CAs are behind human performance.}

The purposes of user intent prediction are threefold. First, it is necessary for CAs to accurately identify user intent in information-seeking conversations. Only in this way can CAs process the information accordingly and use it to provide answers and adjust previous answers. Similar to customer service over phones, routing user questions to different sub-modules in a conversational retrieval system is only possible if the user intent is correctly identified. Second, the CAs need to learn and imitate the behavior of human agents. 
% To illustrate the second point, we use the colorization problem in computer vision. In colorization problems, a neural model is provided with gray scale images as training data and the corresponding colored images as groundtruth. As the models are learning how to colorize different items, it achieves high-level visual understanding~\cite{larsson2016learning} \yfcomment{The colorization problem does not seem to be very similar to our task}.
By identifying user intent in information-seeking conversations, we expect the CA to learn the use of different intent and when to issue requests for more information or details spontaneously. Finally, user intent prediction models can be used to automatically annotate more dialog utterances for data analysis and other tasks such as conversational answer finding.

% \lycomment{The previous passages presented ``Why the task is useful and important''. But it did not cover ``What are related existing works and what are the weakness of existing related studies on user conversation intent''. We should add one more paragraph here to analyze the weakness of existing studies. For example, the user intent modeling in DSTC challenge requires fine-grained user intent labeling. The annotated data is also small and most proposed models in DSTC challenge need feature engineering and are not end-to-end trainable.}

Previous work typically focused on dialog act classification for open-domain conversations~\cite{Stolcke2000Dialogue, LiuEMNLP17, Khanpour2016DialogueAC}. In human-computer chitchat, the goal of the CA is to generate responses that are as realistic as possible with the primary purpose of entertaining. In contrast to chatting, users initiate information-seeking conversations for specific information needs. Human behaviors in chatting and information-seeking conversations can be very different due to the fundamentally distinct purposes. In addition, the Dialog State Tracking Challenges (DSTC)\footnote{\url{https://www.microsoft.com/en-us/research/event/dialog-state-tracking-challenge/}} focus on goal oriented conversations. These tasks are typically tackled with slot filling~\cite{Zhang:2016:JMI:3060832.3061040, DBLP:conf/aaai/YanDCZZL17}. In information-seeking conversations, slot filling is not suitable because of the diversity of information needs. User intent analysis and prediction are needed for an information-seeking setting.

We conduct experiments using two different approaches to predict user intent in information-seeking conversations. Firstly, we extract rich features to capture the content, structural, and sentiment 	
characteristics of utterances and learn models with traditional machine learning (ML) methods. Secondly, we use the implicit representation learning in neural architectures to predict user intent without feature engineering. We then incorporate context information into neural models for enhanced performance. 

\if0
Our contributions can be summarized as follows:

\textbf{(1) We created a conversation dataset called \textit{MSDialog}.} It consists of multi-turn information-seeking dialogs in the technical support domain. We also annotated utterance level user intent in conversations with MTurk. We will release this data with annotated user intent labels to the research community. 

\textbf{(2) We identified the key factors in user intent prediction.} We designed and extracted rich features to predict user intent in information-seeking conversations. These features can be classified into three groups: content, structural, and sentiment. We performed in-depth feature importance analysis on both group level and individual level. 

\textbf{(3) We designed enhanced neural classifiers to predict user intent without explicit feature engineering.} We conducted experiments with several variations of neural models with different incorporated information. We proved that neural models can achieve comparable performances without feature engineering. Moreover, neural models achieved statistically significant improvements over traditional methods after incorporating context information.
\fi

Our contributions can be summarized as follows. (1) We extract rich features including feature groups related to content, structures, and sentiment to predict user intent in information-seeking conversations. We perform an in-depth feature importance analysis on both group and individual level to identify the key factors in this task. (2) We design several variations of neural classifiers to predict user intent without explicit feature engineering. We show that neural models can achieve comparable performance compared to feature engineering based methods. Moreover, neural models achieve statistically significant improvements over traditional methods after incorporating context information. (3) Our experiments show that the trained model achieves good generalization performance on another open benchmark information-seeking conversation dataset (UDC). The code of the implemented user intent prediction models will be released to the research community.\footnote{\url{https://github.com/prdwb/UserIntentPrediction}}

%4.Roadmap: the structure of the whole paper
%\textbf{Roadmap}. 
% \lycomment{The following passage can be commented if there is not enough space left.}
The rest of our paper is organized as follows. In Section~\ref{sec:relatedwork}, we present related work regarding utterance type classification, conversational search, and multi-turn question answering. In Section~\ref{sec:task-and-data}, we formulate the research question of user intent prediction in information-seeking conversations. We also describe the data creation and annotating process. In Section~\ref{sec:baselines}, we extract rich features and learn traditional ML models for user intent prediction. We also perform feature importance analysis in this section. In Section~\ref{sec:neural-models}, we introduce various enhanced neural classifiers for user intent prediction. Section~\ref{sec:conclusion} presents the conclusion and future work.

\section{Related Work}
\label{sec:relatedwork}
%----------------------------------------------------------------------------------------
% \begin{itemize}
%   \item How do existing methods differ from one another and what are their 
%   respective strengths and weaknesses.
%   \item How is this work challenging?
%   \item Create context of the work.
%   \item Create expectation of what the contribution of paper should be.
% \end{itemize}

% Describe related work and the differences between our work with their work

%%----------------------------------------------------------------------------------------
Our work is closely related to utterance classification, conversational search, and multi-turn question answering.

\textbf{Utterance Classification}. Utterance classification is well studied in the NLP and IR domain. Many different classification techniques such as
statistical approaches~\cite{Stolcke2000Dialogue}, SVM, or Hidden Markov Models~\cite{Surendran2006Dialog}
have been used for different applications including
human-human chatting~\cite{Stolcke2000Dialogue},
student's utterance~\cite{Olney:2003:UCA:1118894.1118895}, or forum post classifications~\cite{DBLP:conf/webdb/BhatiaM12}.
However, recent advances in deep learning allow us to use neural architectures for utterance classification both on the word~\cite{DBLP:conf/emnlp/Kim14} and character level~\cite{DBLP:conf/nips/ZhangZL15}. 
These new deep learning techniques have been applied in medical dialog systems~\cite{DBLP:conf/iva/DattaBOWB16}. In this paper, we focus on user intent prediction in information-seeking conversations. This specific utterance classification task presents unique challenges because of the complexity and diversity of human information-seeking conversations.

% \lycomment{The difference between our work and their work is that ...}

% \lycomment{In the related work section, you should not only introduce others' work, but also point out the difference between our work and others' work.}

\ignore{\jtcomment{This section may need to be revised and differently structured. I'll try to come back to this later to help rewrite it. How are DA related to user intent?}}

%%----------------------------------------------------------------------------------------
\textbf{Conversational Search}. Searching via conversational interactions with IR systems is an increasingly popular research topic in both industry and academia~\cite{Zhang2018Rec, Shiga2017modelling, Croft:1987:IRN:35053.35054}. 
% However, the basic concept dates back to earlier work.
\citet{Oddy1977Information} introduced man-machine IR through dialogs without explicitly formulating queries. \citet{belkin1995merit} modeled the human-computer interaction in information-seeking as dialogs. Moreover, information-seeking via conversations is especially important in exploratory search~\cite{Marchionini:2006:ESF:1121949.1121979, DBLP:series/synthesis/2009White}, where users are unfamiliar with the domain they are searching and would rely on effective interactions with retrieval systems. More recently, \citet{Radlinski:2017:TFC:3020165.3020183} identified key properties in conversational IR systems. \citet{Trippas:2018:IDS:3176349.3176387} conducted lab-based observational studies on conversational search and identified that it is more interactive than traditional search and new information-seeking models are needed. In our work, we focus on an essential study in conversational search, which is to predict user intent in this information-seeking setting. Our findings can help build conversational search systems that can provide enhanced searching experience using predicted user intent. 

% \lycomment{The difference between our work and their work is that ...}
%----------------------------------------------------------------------------------------

\textbf{Multi-turn Question Answering}. Early research on multi-turn question answering dates back to AutoTutor~\cite{Graesser2001AutoTutor}, which can simulate human tutors to assist college students to learn computer science. Recent work has focused on single turn QA on factoid questions~\cite{Yin2015Neural} and other open-domain questions (e.g. WikiQA~\cite{Yang2015WikiQA}). The Ubuntu Dialog Corpus~\cite{Lowe2015The} and MSDialog~\cite{Qu2018UserIntent} provide large scale multi-turn QA dialogs in the technical support domain. \citet{DBLP:conf/acl/WuWXZL17} and \citet{Yang2018RespRank} used these datasets to perform conversation response ranking for non-factoid questions. Our work focuses on a specific research need for multi-turn QA in the information-seeking setting. The performance of multi-turn QA could be improved if the user intent is correctly identified. 
% \lycomment{The difference between our work and their work is that ...}
% \todo{update msdialog/RespRanking citation}
% \lycomment{Should add more references on multi-turn QA here refer to \url{ https://scholar.google.com/scholar?start=0&q=Multi-Turn+Question+Answering&hl=en&as_sdt=0,22}}
\section{Task Definition and Dataset}
\label{sec:task-and-data}
%---------------------------------------------
\subsection{Task Definition}
\label{subsec:task}
%---------------------------------------------
The research problem of user intent prediction in information-seeking conversations is defined as follows. The input of the system is an information-seeking dialog dataset $\mathcal{D} = \{(\mathcal{U}_i, \mathcal{Y}_i)\}_{i=1}^N$ and a set of user intent labels $ \mathcal{L} = \{l_1, l_2, \dots, l_c\} $. A dialog $ \mathcal{U}_i = \{u_i^1, u_i^2, \dots, u_i^{k}\} $ contains multiple turns of utterances. $u_i^{k}$ is the utterance at the $k$-th turn of the  $i$-th dialog. $\mathcal{Y}_i$ consists of annotated user intent labels $ \{\mathbf{y}_i^1, \mathbf{y}_i^2, \dots, \mathbf{y}_{i}^k\} $, where $\mathbf{y}_i^k = \{y_i^{k(1)}, y_i^{k(2)}, \dots, y_i^{k(c)}\}$. Here $y_i^{k(m)}, \dots, y_i^{k(n)} = 1$ denotes that the utterance $u_i^{k}$ in dialog $\mathcal{U}_i$ is labeled with user intent \{$l_m, \dots, l_n $\}. Given an utterance $u_i^{k}$ and other utterances in dialog $\mathcal{U}_i$, the goal is to predict the user intent $\mathcal{Y}_i$ of this utterance. The challenge of this task lies in the complexity and diversity of human information-seeking conversations, where one utterance often expresses multiple user intent~\cite{Trippas:2018:IDS:3176349.3176387}.
%---------------------------------------------------
\subsection{Dataset}
\label{subsec:data}
%------------------------------------------
% \lycomment{I reordered the sentences in the following three passages to make it follow a better logic.}

We use the MSDialog dataset that consists of technical support dialogs for Microsoft products. 
%It was crawled from Microsoft Community\footnote{\url{https://answers.microsoft.com}}, which is a well-moderated forum that contains high-quality technical support dialogs between users and agents. 
The data was crawled from the well-moderated Microsoft Community forum\footnote{\url{https://answers.microsoft.com}} which contains high-quality technical support dialogs between users and agents.
The agents include Microsoft staff, community moderators, MVPs, or other experienced product users. Very rich structured data were collected with the dialogs, including the question popularity, answer vote, affiliation of information providers, etc. We choose this dataset because of its information-seeking nature and the well annotated user intent. 
% Dialog datasets with such properties are limited due to privacy issues.
% Although MSDialog has limitations such as narrowed domain and forum-style language, it should be a first step to predict user intent in an information-seeking setting.
% \jtcomment{the above sentence has two concepts, 1, there are not such datasets, 2 this is the first step in predicting user intent.}
Although MSDialog has limitations such as a narrow subject domain and forum-style language, no other openly available dialog datasets with the same detailed annotation exist. We believe this should be a first step to predict user intent in an information-seeking setting.

The dataset contains two sets, a complete set that consists of all the crawled dialogs and a labeled subset that contains only dialogs with user intent annotation. A taxonomy of 12 labels presented in Table~\ref{tab:taxonomy} were developed in~\citet{Qu2018UserIntent} based on work by ~\citet{DBLP:conf/webdb/BhatiaM12} to characterize the user intent in information-seeking conversations. We also present the user intent distribution in Table~\ref{tab:taxonomy}. The complete set consists of 35,000 multi-turn QA dialogs in the technical support domain. Over 2,000 dialogs were selected for user intent annotation on MTurk with the following criteria: (1) With 3 to 10 turns. (2) With 2 to 4 participants. (3) With at least one correct answer selected by the community. (4) Falls into one of the categories of following major Microsoft products: Windows, Office, Bing, and Skype. The inter-rater agreement score was used to ensure the annotation quality. One utterance can be labeled with more than one user intent. A comparison of statistics between the complete set and the labeled subset is presented in Table~\ref{tab:compare-stats}.

\begin{table}[ht]
\centering
\footnotesize
\caption{User intent taxonomy and distribution in MSDialog}
\label{tab:taxonomy}
\vspace{-0.3cm}
\begin{tabular}{llll}
\toprule
\hspace{-0.08in} Code & \hspace{-0.1in} Label \hspace{-0.1in} & Description \hspace{-0.1in}                              \hspace{-0.1in} & \%              \\ \hline
\hspace{-0.08in} OQ   & \hspace{-0.18in} Original Question     \hspace{-0.15in} & The first question from the user to initiate the dialog. \hspace{-0.15in} & 13              \\
\hspace{-0.08in} RQ   & \hspace{-0.18in} Repeat Question       \hspace{-0.15in} & Other users repeat a previous question.                  \hspace{-0.15in} & 3                 \\
\hspace{-0.08in} CQ   & \hspace{-0.18in} Clarifying Question   \hspace{-0.15in} & User or agent asks for clarifications.                   \hspace{-0.15in} & 4                \\
\hspace{-0.08in} FD   & \hspace{-0.18in} Further Details       \hspace{-0.15in} & User or agent provides more details.                     \hspace{-0.15in} & 14                  \\
\hspace{-0.08in} FQ   & \hspace{-0.18in} Follow Up Question    \hspace{-0.15in} & User asks for follow up questions about relevant issues. \hspace{-0.15in} & 5                \\
\hspace{-0.08in} IR   & \hspace{-0.18in} Information Request   \hspace{-0.15in} & Agent asks for information from users.                   \hspace{-0.15in} & 6                 \\
\hspace{-0.08in} PA   & \hspace{-0.18in} Potential Answer      \hspace{-0.15in} & A potential answer or solution provided by agents.       \hspace{-0.15in} & 22              \\
\hspace{-0.08in} PF   & \hspace{-0.18in} Positive Feedback     \hspace{-0.15in} & User provides positive feedback for working solutions.   \hspace{-0.15in} & 6             \\
\hspace{-0.08in} NF   & \hspace{-0.18in} Negative Feedback     \hspace{-0.15in} & User provides negative feedback for useless solutions.   \hspace{-0.15in} & 4     \\
\hspace{-0.08in} GG   & \hspace{-0.18in} Greetings/Gratitude   \hspace{-0.15in} & Greetings or expressing gratitude.                       \hspace{-0.15in} & 22               \\
\hspace{-0.08in} JK   & \hspace{-0.18in} Junk                  \hspace{-0.15in} & No useful information in the utterance.                 \hspace{-0.15in}  & 1           \\
\hspace{-0.08in} O    & \hspace{-0.18in} Others                \hspace{-0.15in} & Utterances cannot be categorized using other classes.    \hspace{-0.15in} & 1            \\ \bottomrule
\end{tabular}
\end{table}

% \lycomment{The following passage should be moved to the experimental setting section instead of being in this data section. We haven't introduce neural models yet. The reviewers will be confused when they read this passage here.}

% All baseline and neural model experiments are supervised and learned with the labeled subset. We also use the complete set to pre-train word embeddings, which is proven to have better performance than using other pre-trained global vectors (GloVe~\cite{pennington2014glove}). 

\begin{table}[ht]
\footnotesize
\centering
\caption{Statistics of MSDialog (complete \& labeled subset)}
\label{tab:compare-stats}
\vspace{-0.3cm}
\begin{tabular}{@{}lll@{}}
\toprule
Items                       & Complete set        & Labeled subset \\ \midrule
\# Dialogs                  & 35,000              & 2,199    \\
\# Utterances               & 300,000             & 10,020   \\
\# Words (in total)         & 24,000,000          & 653,000  \\
Avg. \# Participants        & 3.18                & 2.79     \\
Avg. \# Turns Per Dialog    & 8.94                & 4.56     \\
Avg. \# Words Per Utterance & 75.91               & 65.16    \\ 
\bottomrule
\end{tabular}
\end{table}

% \lycomment{The following passage should be moved to Section 5.3 on generalization on Ubuntu dialogs. Or you can keep it here and refer to Section 5.3 in this passage.}

In order to test the generalization performance of our findings, we use a small portion of UDC that is annotated with the same user intent types. This dataset also consists of multi-turn information-seeking conversations in a technical support domain between an information seeker and provider. However, UDC is generated from internet relay chat (IRC) and contains a significant amount of typos, Internet language, and abbreviations. In addition, UDC contains shorter utterances and more turns per dialog. This part of experiment is presented in Section~\ref{subsec:udc}.

% Consecutive utterances from the same actor are merged as a single utterance. 159 dialogs with 3,185 utterances are sampled to be annotated on an utterance level with the same MTurk protocol as MSDialog.
% \jtcomment{Is UDC a conversation between two actors?}
%\lycomment{One of the SIGIR reviewers mentions}
% \jtcomment{Is it worth mentioning you labeled them? There is a small reference to the labeling in the introduction, would that be sufficient?}

%----------------------------------------------
\subsection{Data Preprocessing}
\label{subsec:data-transformation}
%----------------------------------------------

% \lycomment{Try to remove verbose and duplicate sentences.}

%The user intent prediction task described in Section~\ref{subsec:task} is a multi-label classification problem. For each utterance, there can be more than one user intent label. Thus, we preprocess the data so that it is more suitable for our task.

%We found that the label \textit{Greetings/Gratitude} often co-occurs with other user intent in our previous work~\todo{cite}. 
The purpose of this classification task is to identify and predict user intent so that CAs can process the information accordingly to satisfy the users' information needs. However, utterances which were labeled \textit{Greetings/Gratitude}, \textit{Junk}, and \textit{Others} do not contribute to the purpose of providing information about QA related user intent. Therefore, we remove occurrences of these labels. Note that we only remove these labels if there are more than one label of the given utterance. For example, if the annotation for the given utterance is \textit{GG+OQ}, we transform the annotation into \textit{OQ}. If the annotation is just \textit{GG}, no transformation is needed. This reduces the number of unique label combinations from 316 to 152. 

\begin{figure}[h!]
  \begin{subfigure}[t]{.23\textwidth}
  \centering
    \includegraphics[width=.9\textwidth]{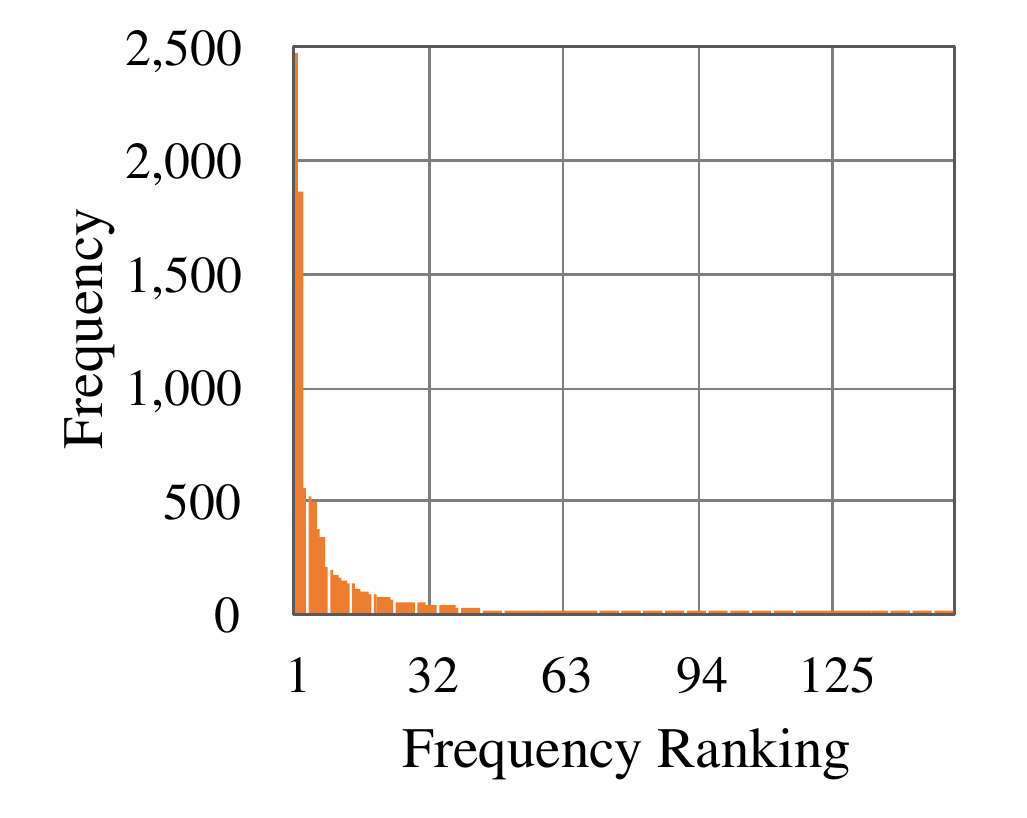}
    \caption{Label combination occurrence from the most frequent to the least frequent.}
    \label{fig:label-dist}
  \end{subfigure}\hfill
  \begin{subfigure}[t]{.23\textwidth}
  \centering
    \includegraphics[width=.9\textwidth]{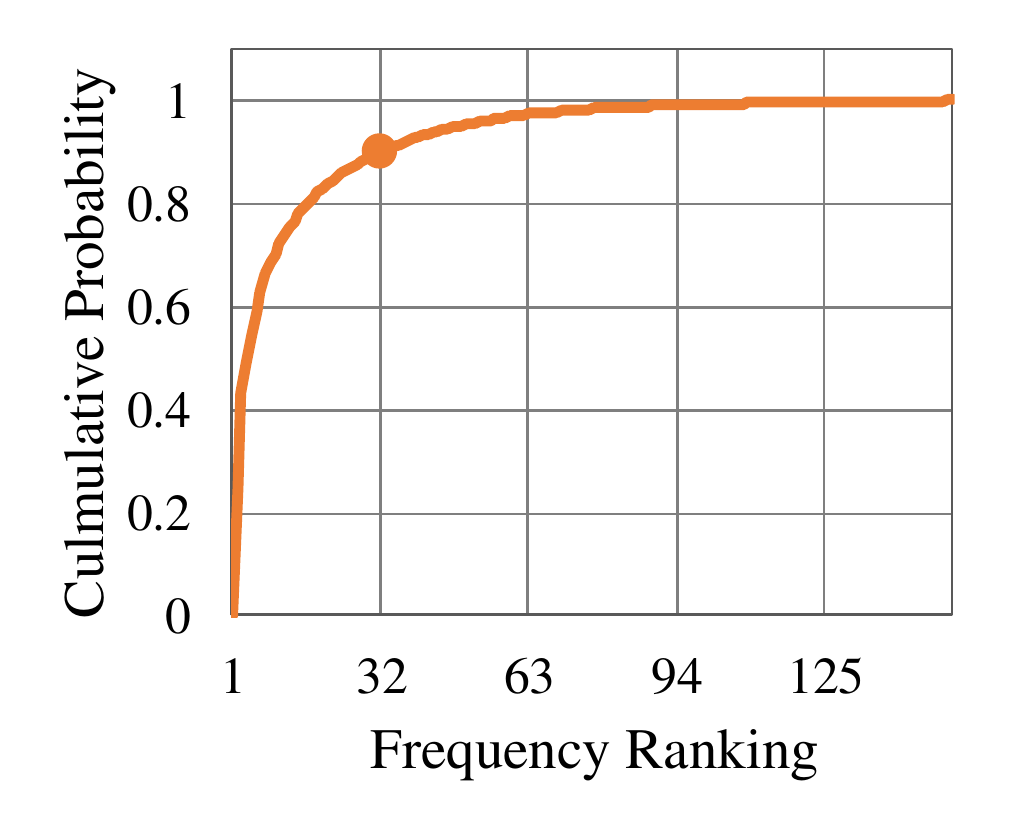}
    \caption{Cumulative occurrence probability added from the most frequent to the least frequent. The marked point is $(32, 0.9)$.}
    \label{fig:label-dist-incre}
  \end{subfigure}
  \vspace{-0.3cm}
  \caption{Label combination distribution}
\end{figure}

In addition, some label combinations of user intent labels are quite rare in the data. As indicated in Figure~\ref{fig:label-dist}, the top frequent label combinations have hundreds of occurrences in the data (e.g. \textit{PA}, \textit{OQ}, \textit{PF}, \textit{FD+PA}, \textit{FD}), while the least frequent labels only have exactly one occurrence (e.g. \textit{CQ+FD+IR+RQ}, \textit{CQ+FD+FQ+PF}). These rare label combinations are very likely due to minor annotation errors or noise with MTurk. Annotation quality assurance was performed based on the dialog-level inter-rater agreement~\cite{Qu2018UserIntent} to keep the complete dialog intact and thus may result in minor noise on an utterance level. We also plot the cumulative distribution of label combinations for better illustration in Figure~\ref{fig:label-dist-incre}. The most frequent 32 label combinations constitute 90\% of total label combination occurrences as marked in the figure. All 12 user intent labels are individually present in these 32 most frequent combinations except for \textit{Others}. For the rest of the label combinations, we randomly sample one of the labels from each combination as the user intent label for the given utterance. For example, if the annotation for the given utterance is \textit{CQ+FD+IR+RQ}, we transform it into a single label by randomly sampling one of the four labels, such as \textit{CQ}. Therefore, the total number of label combinations in the data was reduced to 33 (including \textit{Others}). 
%\lycomment{In this way, we transferred the multi-label classification setting to a single-label classification setting with minimal change of user intent annotations. }\cqcomment{it is still  multi-label classification}. \cqcomment{In this way, we ensure the homogeneity of the data with minimal change to user intent annotations.}
We adopted this setting since these rare label combinations are very likely due to minor annotation errors. 
In addition, it would be very difficult to learn a prediction model for these label combinations with few instances.

% The preprocessing steps to the data has little impact on labeling quality as we only remove labels from very rare label combinations. This steps can help reduce the noise in the data and be beneficial to model training. Finally, we only use the first 100 tokens of the utterances.
% \jtcomment{Which label combinations did you discard? Why only the first 100 tokens?}
% \cqcomment{Using only 100 tokens is a major design flaw, I am fixing it. Thank you pointing this out!}\lycomment{Yes, ''only use first 100 tokens'' should be fixed. We can maintain the original length during the data reprocessing step. The ''maxUtteranceLength'' can be a hyper-parameter of the neural models. We can do the padding/length reducing when we generate the batches for training.  ''maxUtteranceLength'' can be optimized using the validation data.}

For UDC, we observe a similar label combination distribution. So we preprocess the data in the same way. We have 34 label combinations for the UDC with 27 of them overlapping with MSDialog.
\section{Baselines and Feature Analysis}
\label{sec:baselines}
%----------------------------------------------------------------
In this section, we extract several features following previous work~\cite{DBLP:conf/webdb/BhatiaM12, DBLP:conf/acl/DingCLZ08} and adopt different ML methods to build baseline models. In addition to reporting baseline performance, we also perform feature importance analysis to identify key factors in user intent prediction.
% \jtcomment{the above section is confusing because you use ''training set'', ''validation set'' and ''test set'' while in previous sections you talk about the ''whole set'' and ''labeled subset''.}
% \lycomment{You can add a data statistics table for the train/valid/test data. You can refer to Table 2 of this SMN paper \url{http://www.aclweb.org/anthology/P17-1046. }. You can also add more descriptions on how to split the labeled subset into train/valid/test partitions.}

\subsection{Features}
\label{subsec:features}
\begin{table*}[ht]
	\centering
	\footnotesize
		\caption{Features extracted for user intent prediction in information-seeking conversations.}
		\label{tab:features}
		\vspace{-0.3cm}
		\begin{tabular}{ l l l l }
			\toprule
			Feature Name	 & Group & Description & Type \\
			\midrule
			Initial Utterance Similarity     & Content     & Cosine similarity between the utterance and the first utterance of the dialog  & Real   \\
			Dialog Similarity          & Content     & Cosine similarity between the utterance and the entire dialog                  & Real   \\ 
			Question Mark              & Content     & Does the utterance contain a \textit{question mark}                                      & Binary \\
			Duplicate                  & Content     & Does the utterance contain the keywords \textit{same}, \textit{similar}                           & Binary \\
			5W1H                        & Content     & Does the utterance contain the keywords \textit{what}, \textit{where}, \textit{when}, \textit{why}, \textit{who}, \textit{how}        & One-hot vector  \\ \hline
			Absolute Position          & Structural  & Absolute position of an utterance in the dialog                            & Numerical \\
			Normalized Position       & Structural  & Normalized position of an utterance in the dialog (AbsPos divided by \# utterances)  & Real\\
			Utterance Length              & Structural  & Total number of words in an utterance after stop words removal             & Numerical \\
			Utterance Length Unique        & Structural  & Unique number of words in an utterance after stop words removal            & Numerical \\
			Utterance Length Stemmed Unique & Structural  & Unique number of words in an utterance after stop words removal and stemming & Numerical \\ 
			Is Starter                & Structural  & Is the utterance made by the dialog starter                                & Binary\\ \hline
			Thank                       & Sentiment   & Does the utterance contain the keyword \textit{thank}                               & Binary \\ 
			Exclamation Mark           & Sentiment   & Does the utterance contain an \textit{exclamation mark}                             & Binary \\
			Feedback             & Sentiment   & Does the utterance contain the keyword \textit{did not}, \textit{does not}                  & Binary \\
			Sentiment Scores           & Sentiment   & Sentiment scores of the utterance computed by VADER~\cite{DBLP:conf/icwsm/HuttoG14} (positive, neutral, and negative)  & Real \\
			Opinion Lexicon            & Sentiment   & Number of positive and negative words from an opinion lexicon              & Numerical \\
			\bottomrule
		\end{tabular}

\end{table*}

We extract three groups of features to detect user intent in information-seeking conversations, including content, structural, and sentiment features. An overview of the features is provided in Table~\ref{tab:features}. Although MSDialog is derived from forum data, it is considered as a dialog dataset~\cite{Qu2018UserIntent}. Thus, we refrain from developing features that can only be extracted from the metadata of the forum, such as user authority level or answer votes, so that our method can be applicable to dialog systems. The features are designed to capture the content and sentiment characteristics of the utterances as well as the structural information of the dialogs. 

\textbf{Content features}. We build a TF-IDF representation of utterances and compute the cosine similarity of the given utterance with the dialog initial utterance (which typically is the question that initiates the QA dialog), and the entire dialog. These features are meant to capture the relevance level of the given utterance to the dialog in a general way. In addition, the presence of question marks is a strong indicator that the current utterance contains a question. Moreover, we assume that 5W1H keywords (what, where, when, why, who, and how) can suggest the type of the question.

\textbf{Structural features}. The position of an utterance in a dialog can reveal crucial information about user intent. Intuitively, answers tend to be at even number positions in a dialog, while user feedback and follow up questions tend to be at odd number positions. In addition, we include the utterance length with and without duplication removal and stemming. We analyzed the data and found that utterances containing positive feedback are relatively short, while utterances containing questions or answers tend to be long as they typically contain details. Finally, if the given utterance is generated by the information seeker (dialog starter), it is more likely to contain user related questions or user feedback. The structural features not only provide individual characteristics for utterances, but also evaluate the utterances on a dialog level.

\textbf{Sentiment features}. We expect sentiment features to be useful in identifying user feedback and gratitude expressions. In information-seeking conversations, positive and negative sentiments do not necessarily determine the feedback type. However, we expect them to be correlated to some extent. We also include classic indicators of sentiment, such as the presence of ``thank'', ``does not/did not'' and exclamation marks. In addition, we use VADER~\cite{DBLP:conf/icwsm/HuttoG14} to compute the positive/negative/neutral sentiment scores. We also count the number of positive and negative words using an opinion lexicon~\cite{Liu:2005:OOA:1060745.1060797}.

%\lycomment{mark 1752 2018-0502}

%---------------------------------------------------------
\subsection{Methods and Evaluation Metrics}
\label{subsec:baseline-methods}
%---------------------------------------------------------

\subsubsection{Methods}
For each utterance, we extract a set of features as described in Section \ref{subsec:features}. To apply traditional ML methods with features to this task, we need to transform this multi-label classification problem to multi-class classification. Three transformation strategies are typically used: binary relevance, classifier chains, and label powerset. Binary relevance does not consider the label correlations and label powerset generates new labels for every label combination. So we choose classifier chains as the transformation strategy for traditional ML methods. This strategy performs binary classifications for each label and take predictions for previous labels as extra features. This transformation strategy is the best fit for our task as it considers the label dependency without explicitly generating new labels for every label combination. We adopt classic ML methods, including Naive Bayes classifier, SVM, random forest, and AdaBoost as baseline classifiers. In addition, we use ML-kNN, which supports multi-label classification by nature.

% \cqcomment{Should we remove/ shorten the following part?}
% Binary relevance produces binary classification results for every label and does not take the label correlations into consideration. Thus this method is not suitable for this task as clear label co-occurrence patterns exist in the data\todo{cite}. Label powerset considers each different combination of labels in the training set as a single label. Due to the data transformation performed in Section~\ref{subsec:data-transformation}, our data could favor this strategy. However, label powerset does not generalize well because it cannot predict unseen label combinations in held out data. 

% \jtcomment{It may be useful to change ''machine learning'' to ''ML'' to save some space.}

\subsubsection{Metrics}
Due to the nature of the intent prediction task, we adopt metrics suitable for multi-label classification problems.

\textbf{Accuracy} (Acc). It is known as the intersection over the union (IoU) in the multi-label classification settings. Accuracy is defined as the number of correctly predicted labels divided by the union of predicted and true labels for every utterance. For example, if the model predicts ``\textit{PA+CQ}'' for the given utterance while the ground truth is ``\textit{PA+IR}'', then the accuracy is $\frac{1}{3}$. The reported performance is the average metric over all utterances.
% \lycomment{Why does label-wise accuracy leverage precision and recall ?}

\textbf{Precision, recall and F1 score}. Precision is defined as the number of correctly predicted labels divided by predicted labels. Recall is defined as the number of correctly predicted labels divided by true labels. F1 is their harmonic mean. These metrics provide an overall performance evaluation for all utterances.

%----------------------------------------------------------
\subsection{Main Experiments and  Results}
\label{subsec:baseline-exp}
%------------------------------------------------------------
\subsubsection{Experimental Setup}
\label{subsubsec:baseline-exp-setup}
%------------------------------------
We split the labeled subset of MSDialog into training, validation, and test sets. Table~\ref{tab:set-stats} gives the statistics of the three sets. We have to point out that although this dataset is not that large, the annotation cost for such fine-grained user intent in conversations is quite high (estimated around \$1,700 on MTurk according to~\cite{Qu2018UserIntent}). This dataset size is also larger than the data used in related work~\cite{DBLP:conf/webdb/BhatiaM12, Surendran2006Dialog}. The models are trained with scikit-multilearn\footnote{\url{http://scikit.ml/}} and scikit-learn\footnote{\url{http://scikit-learn.org/stable/}} on the training set. We tune the hyper-parameters on the validation set based on accuracy and report the performance on the test set. %The tuned hyper-parameters include the number of estimators and stopping criteria for random forest, etc.

\begin{table}[htbp]
\footnotesize
\centering
\caption{Statistics of training, validation, and testing sets}
\label{tab:set-stats}
\vspace{-0.3cm}
\begin{tabular}{@{}llll@{}}
\toprule
Item                        & Train & Val & Test  \\ \midrule
%\# Dialogs                  & 1,760 & 220        & 219   \\
\# Utterances               & 8,064 & 986        & 970   \\
Min. \# Turns Per Dialog    & 3     & 3          & 3     \\
Max. \# Turns Per Dialog    & 10    & 10         & 10    \\
Avg. \# Turns Per Dialog    & 4.58  & 4.48       & 4.43  \\
Avg. \# Words Per Utterance & 70.42 & 67.53      & 68.64 \\ \bottomrule
\end{tabular}
\end{table}
%------------------------------------
\subsubsection{Baseline Results}
\label{subsubsec:baselien-results}
%-------------------------------------

The baseline results are presented in Table~\ref{tab:baseline-results}. Two ensemble methods, random forest and AdaBoost achieve the best overall performance of all baseline classifiers. AdaBoost achieves the best accuracy while random forest achieves the best F1\ignore{\jtcomment{check consistency of spelling ''F1'' or ''F-1''}} score. Surprisingly, ML-kNN performs relatively poorly despite its nature of an adapted algorithm for multi-label classification.

\begin{table}[htbp]
\footnotesize
\centering
\caption{Experiment results for baseline classifiers}
\label{tab:baseline-results}
\vspace{-0.3cm}
\begin{tabular}{@{}lllll@{}}
\toprule
Methods    & Acc & Precision       & Recall          & F1              \\ \midrule
ML-kNN     & 0.4715                & 0.6322          & 0.4471          & 0.5238          \\ 
NaiveBayes & 0.4870                & 0.5563          & 0.4988          & 0.5260          \\
SVM        & 0.6342                & 0.7270          & 0.5847          & 0.6481          \\
RandForest & 0.6268                & \textbf{0.7657} & 0.5903          & \textbf{0.6667} \\
AdaBoost   & \textbf{0.6399}       & 0.7247          & \textbf{0.6030} & 0.6583          \\\bottomrule
\end{tabular}
\end{table}

%---------------------------------------------------------

\subsection{Additional Feature Importance Analysis}
\label{subsec:feature-importance}
%---------------------------------------------------------

\subsubsection{Feature Group Analysis}
\label{subsubsec:feature-group}
%--------------------------------------------------------
We use one of the best baseline classifiers, random forest, and different combinations of feature groups to analyze the feature importance on a group level. The hyper-parameters are set to the best ones tuned on all features .

% \begin{table}[h]
% \footnotesize
% \centering
% \caption{Results of random forest with different groups of features}
% \label{tab:feature-group}
% \begin{tabular}{lllll}
% \toprule
% Feature Group(s)       & Accuracy & Precision       & Recall          & F1             \\ \midrule
% Content                & 0.5272                                                          & 0.6097          & 0.4821          & 0.5384          \\
% Structural             & \textbf{0.5809}                                                 & \textbf{0.6871} & \textbf{0.5434} & \textbf{0.6068} \\
% Sentiment              & 0.3306                                                          & 0.4087          & 0.3222          & 0.3603          \\ \hline
% Content+Structural   & 0.6081                                                          & 0.7393          & 0.5640          & 0.6399          \\
% Content+Sentiment    & 0.5577                                                          & 0.6523          & 0.5179          & 0.5774          \\
% Structural+Sentiment & \textbf{0.6110}                                                 & \textbf{0.7569} & \textbf{0.5672} & \textbf{0.6485} \\ \hline
% All                    & \textbf{0.6268}                                                 & \textbf{0.7657} & \textbf{0.5903} & \textbf{0.6667} \\ \bottomrule
% \end{tabular}
% \end{table}

For using a single feature group, structural features is the most important feature group as presented in Table~\ref{tab:feature-group}. Structural features and content features are significantly more important than sentiment features. We expect the sentiment features to capture the sentiment in user feedback but they might not be able to effectively discriminate other user intent. Structural features provide better performance than content features.  We believe that this can be explained by the fact that hand-crafted content features cannot capture the complex user intent dynamics in human information-seeking conversations. 
% Although human multi-turn QA conversations follow clear patterns~\cite{Qu2018UserIntent}, information gained from structural patterns should not be better than the content itself.
% This claim is too strong. It is possible that the structure information is more useful.

\begin{table}[htbp]
\footnotesize
\centering
\caption{Experiment results for different feature groups}
\label{tab:feature-group}
\vspace{-0.3cm}
\begin{tabular}{lllll}
\toprule
Group(s)       & Acc & Precision       & Recall          & F1             \\ \midrule
Content                & 0.5272                                                          & 0.6097          & 0.4821          & 0.5384          \\
Structural             & \textbf{0.5809}                                                 & \textbf{0.6871} & \textbf{0.5434} & \textbf{0.6068} \\
Sentiment              & 0.3306                                                          & 0.4087          & 0.3222          & 0.3603          \\ \hline
Con+Str   & 0.6081                                                          & 0.7393          & 0.5640          & 0.6399          \\
Con+Sen    & 0.5577                                                          & 0.6523          & 0.5179          & 0.5774          \\
Str+Sent & \textbf{0.6110}                                                 & \textbf{0.7569} & \textbf{0.5672} & \textbf{0.6485} \\ \hline
All                    & \textbf{0.6268}                                                 & \textbf{0.7657} & \textbf{0.5903} & \textbf{0.6667} \\ \bottomrule
\end{tabular}
\end{table}

For combinations of two feature groups, content+structural features and structural+sentiment features achieve comparable results. However, structural+sentiment features achieve slightly higher results on all metrics. The performance of using two groups of features is higher than using one of these two feature groups individually. Thus, combining structural features with another feature group boosts the performance of using structural features alone. Interestingly, content+sentiment features is unable to outperform the structural features alone. The results of using all features is the highest among all settings, confirming that all feature groups contribute to the performance of user intent prediction. 

%---------------------------------------------------------

\subsubsection{Feature Importance Scores}
\label{subsubsec:feature-scores}

In the previous section, we evaluated the feature importance on a group level. In this section we focus on individual features to provide a more fine-grained analysis. We use random forest to output individual feature importance scores.\footnote{\url{https://scikit-learn.org/stable/auto_examples/ensemble/plot_forest_importances.html}} As described in Section~\ref{subsec:baseline-methods}, we used classifier chains to transform this multi-label classification problem. This method expands the feature space by including previous label predictions as new features for the current label prediction. This makes it not appropriate to evaluate original features. Thus, we use the Label Powerset method as the data transformation strategy for this section. The relative feature importance scores are presented in Table~\ref{tab:feature-importance}. This analysis can identify key factors in user intent prediction.

\begin{table}[htbp]
\footnotesize
\centering
\caption{Individual feature importance from a random forest classifier with relative importance scores. ``Str'', ``Con'', ``Sen'' refer to ``Structural'', ``Content'', ``Sentiment'' respectively.} 
\vspace{-0.3cm}
\label{tab:feature-importance}
\begin{tabular}{llll|llll}
\toprule
Rank & Feature                    & Group      & Impt & Rank & Feature               & Group     & Impt \\ \midrule
1    & AbsPos              & Str & 1.0        & 13   & Lex(Pos) & Sen & 0.2814     \\
2    & InitSim    & Con    & 0.9745     & 14   & Lex(Neg) & Sen & 0.2337     \\
3    & NormPos             & Str & 0.8684     & 15   & Thank                      & Sen & 0.1607     \\
4    & Starter                      & Str & 0.8677     & 16   & How                        & Con   & 0.08074    \\
5    & DlgSim               & Con    & 0.6778     & 17   & Dup                  & Con   & 0.06908    \\
6    & SenScr(Neu)      & Sen  & 0.6465     & 18   & What                       & Con   & 0.06576    \\
7    & SenScr(Pos)     & Sen  & 0.5601     & 19   & ExMark           & Sen & 0.06424    \\
8    & Len                & Str & 0.5335     & 20   & When                       & Con   & 0.05989    \\
9    & LenUni        & Str & 0.4381     & 21   & Feedback              & Sen & 0.02859    \\
10   & LenStem & Str & 0.4354     & 22   & Where                      & Con   & 0.02356    \\
11   & SenScr(Neg)     & Sen  & 0.3495     & 23   & Why                        & Con   & 0.0232     \\
12   & QuestMark                    & Con    & 0.3003     & 24   & Who                        & Con   & 0.01423    \\ \bottomrule
\end{tabular}
\end{table}

We summarize our observations as follows:
(1) Structural features including ``Absolute Position'', ``Normalized
Position'', ``Is Starter'' are ranked in the top-5 in terms of feature importance. Moreover, other structural features, such as various forms of utterance length are observed to be relatively informative in general. This confirms the results in Section~\ref{subsubsec:feature-group} that the structural feature group is the most important one.
(2) ``Initial Utterance Similarity'' and ``Dialog Similarity'' are content features that can be highly informative for identifying user intent. Both features are indicators of how closely the utterance connects with the information-seeking process. Other content features, such as ``5W1H'', however, contribute little to predicting user intent.
(3) Some sentiment features are relatively important in identifying user intent, such as positive and neutral sentiment scores. However, some other sentiment features contribute little to the task, such as the existence of exclamation marks and ``thank''.
(4) We observe that features ranked from the 15\textsuperscript{th} to the last one in Table~\ref{tab:feature-importance} are all ``keyword features''. These features are based on a simple rule that whether the given utterance contains pre-defined keywords. For example, the ``5W1H'' feature looks for ``what/where/when/why/who/how'' in the given utterance and the ``Feedback'' feature looks for ``did not/does not''. The major drawback of manual feature engineering is amplified in this task due to the complexity and diversity of human information-seeking conversations.
% \jtcomment{In the Table you have on rank 21 ''have feedback'' is this a ''keyword feature''? If so, it may be useful to provide an example.}

\section{Enhanced Neural Classifiers}
\label{sec:neural-models}
%-------------------------------------------
We expected the content of an utterance to be a good indicator of user intent types compared to other features. \ignore{\lycomment{This is an overclaim. How about structure features like the ``Absolute Position'' ? You can say ``the content and sequential information of utterances is important for user intent type prediction''. or ''to the best of our knowledge...''} }However, as shown in Section~\ref{subsec:feature-importance}, the hand-crafted content features are unable to capture the complex characteristics of human information-seeking conversations. Thus, in this section we adopt neural architectures to automatically learn representations of utterances without feature engineering.

%-------------------------------------------
\subsection{Our Approach}
\label{subsec:our-approach}
%------------------------------------------
\subsubsection{Base Models}
\label{subsubsec:cnn}
%-------------------------------------------
Given the previous success in modeling text sequences using CNN and bidirectional LSTM (BiLSTM)~\cite{GRAVES2005602}, we choose these two architectures as our base models. Although utterances are grouped as dialogs, the base models consider utterances independently.
% \lycomment{change to ``independent sequences'', which could be a group of multiple sentences.}. 

Given an utterance $u_i^{k} = \{w_1, w_2, \dots, w_m\}$ (the $k$-th utterance in the $i$-th dialog) that contains $m$ tokens, we first transform the sequence of tokens into a sequence of token indices $\bm{S} = \{s_1, s_2, \dots, s_m\}$. Then we pad the sequence $\bm{S}$ to a fixed length $n$ (the max sequence length). 
% If $n > m$, the padding procedure adds integer $0$ at the beginning of the sequence as the default setting in Keras. After padding, the sequence $\bm{S}$ becomes $\{0, \dots, 0, s_1, s_2, \dots, s_m\}$ with the sequence length of $n$. If $n < m$, the padding procedure truncates the sequence at the beginning of the sequence to $n$ tokens. \lycomment{If the space is limited, the detailed descriptions of padding process could be removed. Since this is well known in the NN\&NeuIR community.}
Both CNN and BiLSTM start with an embedding layer initiated with pre-trained word embeddings. Preliminary experiments indicated that using MSDialog (complete set) to train word embeddings is more effective than using GloVe~\cite{pennington2014glove} in terms of final model performance. The embedding layer maps each token in the utterances to a word embedding vector with a dimension of $d$.

%The embedding layer is essentially a look-up table that maps a word index to its pre-trained word embedding vector of a fixed dimension of $d$, which is typically set to 100, 200, or 300. After the embedding layer, the input utterance be tuned into a tensor with the shape of $(n, d)$.

We focus on the CNN model following previous work~\cite{DBLP:conf/emnlp/Kim14} here, because it achieves better performance in our experiments. After the embedding layer, filters with the shape $(f, d)$ are applied to a window of $f$ words. $f$ is also referred to as the filter size. Concretely, a convolution operation is denoted as
\begin{footnotesize}
\begin{equation}\label{eqn:conv}
\begin{aligned}
c_i = \sigma(\mathbf{w} \cdot \mathbf{e}_{i:i+f-1} + b)
\end{aligned}
\end{equation}
\end{footnotesize}
Where $c_i$ is the feature generated by the $i$-th filter with weights $\mathbf{w}$ and bias $b$. This filter is applied to an embedding matrix, which is the concatenation from the $i$-th to the $(i+f-1)$-th embedding vectors. An non-linearity function (ReLU) is also applied. This operation is applied to every possible window of words and generates a feature map $\mathbf{c} = \{c_1, c_2, \dots, c_{n-f+1}\}$. More filters are applied to extract features of the utterance content. Max pooling are applied to select the most salient feature of a window of $f'$ features by taking the maximum value $\hat{c_i} = max\{\mathbf{c}_{i:i+f'-1}\}$. $f'$ denotes the max pooling kernel size. A dropout layer is applied after the pooling layer for regularization. %The  %convolution-pooling-dropout layers could be added multiple times. 

% The output is the input for the next convolutions-pooling-dropout layer combination. 

After the last convolutional layer, we perform global max pooling by taking the maximum value $\hat{c} = max\{\mathbf{c}\}$ for the feature map $\mathbf{c}$ at this step. This operation reduces the dimension of the tensor to one. This tensor is further transformed to an output tensor of shape $(1, l)$, where $l$ is the number of user intent labels (12 for our task). Sigmoid activation is applied to each value of the output tensor to squash the value to a confidence level between 0 and 1. A threshold is chosen to determine whether the given label present in the final prediction. If the model is not confident of predicting any label, the label of the highest confidence level is the prediction. We tuned the threshold with the validation data. %We use the input of one utterance for easy illustration, but neural architectures take a batch of utterances as input in practice.

%-------------------------------------------
\subsubsection{Incorporate Context Information}
\label{subsubsec:cnn-context}
%-------------------------------------------

As shown in the previous work~\cite{Qu2018UserIntent}, user intent follows clear flow patterns in information-seeking conversations. The user intent of a given utterance is closely related to the utterances around it, which compose the \textit{context} for the given utterance. Incorporating context information into neural models is easier compared to that for traditional ML methods shown in Section~\ref{sec:baselines}. We consider two ways as follows.

\textbf{Direct Expansion}. The most straightforward way to incorporate context information is to expand the given utterance with its context. Concretely, the expanded utterance for $u_i^k$ is $\hat{u}_i^k = u_i^{k-1} \oplus u_i^{k} \oplus u_i^{k+1} $, where $\oplus$ is the concatenate operator. $\hat{u}_i^k$ is considered as the given utterance in base models.

\begin{figure}[ht]
  \begin{subfigure}[t]{.281\textwidth}
  \centering
    \includegraphics[width=.99\textwidth]{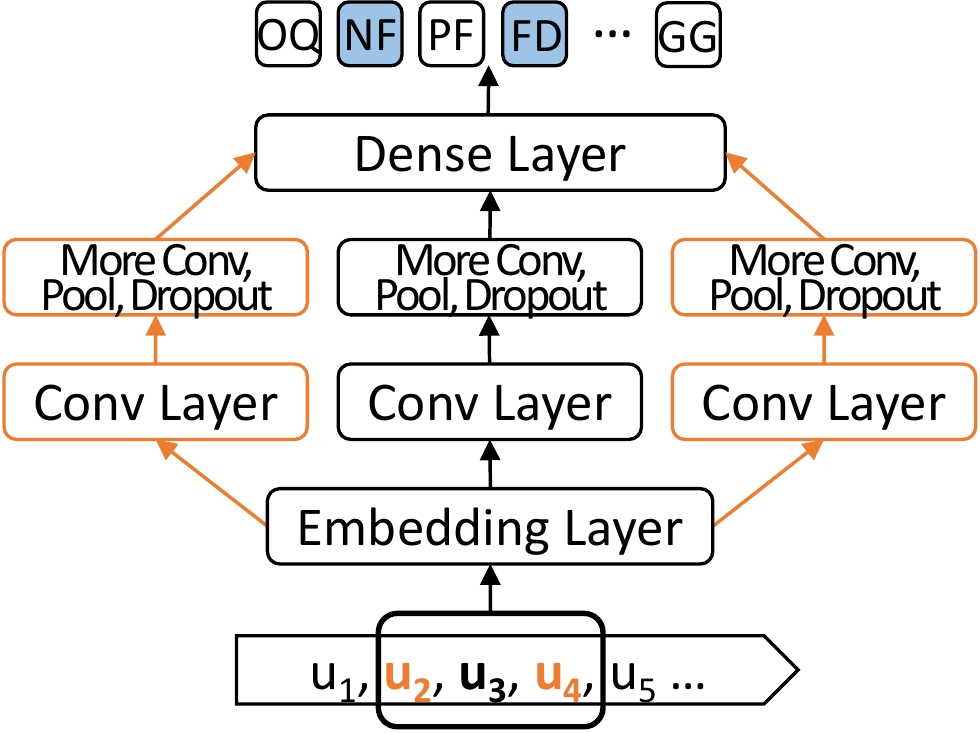}
    \vspace{-0.14in}
    \caption{CNN-Context-Rep}
    \label{fig:cnn-context-rep}
  \end{subfigure}\hfill
  \begin{subfigure}[t]{.189\textwidth}
  \centering
    \includegraphics[width=.99\textwidth]{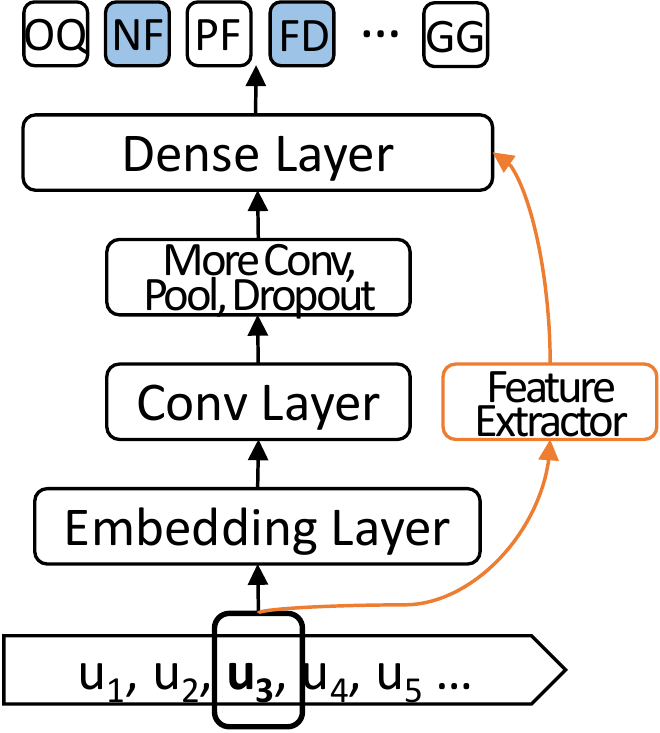}
    \vspace{-0.14in}
    \caption{CNN-Feature}
    \label{fig:cnn-feature}
  \end{subfigure}
%   \vspace{-0.12in}
    \vspace{-0.3cm}
  \caption{Architectures for enhanced neural classifiers. Components marked orange are extra information incorporated into the base CNN model (black). The utterance in bold is the current utterance. Predicted labels are marked blue.}
\end{figure}

\textbf{Context Representation}. Given an expanded utterance $\hat{u}_i^k$ as input, the neural architecture first segments it into three original utterances of $u_i^{k - 1}$, $u_i^{k}$, and $u_i^{k + 1}$. We apply convolution operations and max pooling to the utterances separately as shown in Figure~\ref{fig:cnn-context-rep}. After global pooling following the last convolutional layer, the three one-dimensional tensors are concatenated for final predictions. This approach extracts features from the given utterance and its context separately. Thus, we are able to learn the importance of the given utterance and its context by tuning context-specific hyper-parameters, such as the number of filters for context utterances. 
% \lycomment{The previous sentence is a bit strange. Do you mean 1) the model could learn the importance score of different context utterances or 2) the important context information has been captured by the model ?}

% We use a different approach to incorporate context information. 
% \begin{figure}[h]
%     \centering
%     \includegraphics[width=1.0\linewidth]{img/CNN-Context-Rep_crop.pdf}
%     \caption{Architecture for CNN-Context-Rep \todo{to be updated}}
%     \label{fig:cnn-context-rep}
% \end{figure}

%-------------------------------------------
\subsubsection{Incorporate Extra Features}
\label{subsubsec:cnn-feature}
%------------------------------------------
We found many useful features for user intent prediction such as structural features in the feature importance analysis in Section~\ref{subsec:feature-importance}. % to evaluate which features are informative in predicting user intent. We assume that some of the features can be implicitly captured by neural architectures.
However, nearly half of the features can not be exploited by only looking at a single utterance. For example, normalized/absolute utterance position and utterance similarity with the dialog/initial utterance cannot be captured without a holistic view over the entire dialog. Some of the uncaptured features are highly informative. This motivates us to incorporate hand-crafted features into the neural architectures. As shown in Figure~\ref{fig:cnn-feature}, all hand-crafted features described in Section~\ref{subsec:features} are incorporated into neural architectures at the last dense layer. The feature vector is concatenated with the neural representation of the utterance before making final predictions. %We adjust the importance of hand-crafted features with the utterance representation by tuning the dimension of the utterance representation tensor. 
% \lycomment{Why is the dimension of feature tensor fixed ? It should be a hyper-parameter in the dense layer that can be tuned.} 

Finally, we combine two base models with various extra components to produce several systems for comparison as follows:

\begin{itemize}[leftmargin=*]
    \item \textbf{CNN}. The base CNN model that consists of three convolutional layers with the same filter size.
    % It takes the word embedding of the given utterance as input and predicts one or more user intent.
    \item \textbf{CNN-Feature}. The CNN model that incorporates extra hand-crafted features at the last layer.
    \item \textbf{CNN-Context}. The CNN model that incorporates context information with direct expansion.
    \item \textbf{CNN-Context-Rep}. The CNN model that incorporates context information with context representation.
    \item \textbf{BiLSTM}. The BiLSTM model that represents the given utterance both in the ordinary order and the reverse order. 
    \item \textbf{BiLSTM-Context}. The BiLSTM model that incorporates context information with direct expansion.
\end{itemize}

% \lycomment{It is better to summarize different neural model here to introduced neural models explored in this paper: CNN, BiLSTM, CNN-Context, BiLSTM-Context, CNN-Feature, CNN-MulFilterS, CNN-Context-Rep. For each model, use 1-3 sentence to describe it. Organize them as an itemize list. }

% \lycomment{The description of neural models and the analysis of experimental results are usually separated. So we need to explain clearly what do ``CNN, BiLSTM, CNN-Context, BiLSTM-Context, CNN-Feature, CNN-MulFilterS, CNN-Context-Rep '' mean before we enter into Section 5.2 on experiments and result analysis. }

% \begin{figure}
%     \centering
%     \includegraphics[width=1.0\linewidth]{img/cnn-feature.png}
%     \caption{Architecture for CNN-Feature}
%     \label{fig:cnn-feature}
% \end{figure}
%------------------------------------------
\subsection{Experiments and Evaluation}
\label{subsec:exp}
%-----------------------------------------
\subsubsection{Neural Baselines}
\label{subsubsec:neural-baselines}
%----------------------------------------
In addition to the base model of \textbf{BiLSTM} and \textbf{CNN}, we further introduce two commonly used neural models for text classification as baselines. For both new baselines, we modify the models to generate multi-label predictions.

\textbf{CNN-MFS}. The CNN model with multiple filter sizes as described by~\citet{DBLP:conf/emnlp/Kim14} is a pioneer model to apply neural networks to text classification. This model uses different filter sizes of 3, 4, and 5 to generate feature maps of different window sizes. 
% After global pooling of each feature map, all features are concatenated for the final prediction. 
% Kim conducted experiments with different word embedding initialization methods. We use pre-trained word embeddings to initialize the embedding layers for a consistent setup with our other experiments. 
% \lycomment{Is this the same setting with the word embeddings used by our models ? If yes, we don't have to say this here since the experimental setup already included this.}

\textbf{Char-CNN}. \citet{DBLP:conf/nips/ZhangZL15} introduced a character-level CNN for text classification. There are two variants of the model, a large one and a small one, depending on the numbers of convolutional filters and dense layer units. We report the performance on the small model as it achieves better results in our task.
%----------------------------------------
\subsubsection{Experimental Setup}
\label{subsubsec:exp-setup}
%-----------------------------------------
We use the same data and metrics as in baseline experiments in Section~\ref{sec:baselines}. All models are implemented with TensorFlow\footnote{\url{https://www.tensorflow.org/}} and Keras\footnote{\url{https://keras.io/}}. Hyper-parameters are tuned with the validation data. We found that setting (convolutional filters, dropout rate, dense layer units, max sequence length, convolutional filters for context, and dense layer units for context) to (1024, 0.6, 256, 800, 128, 128) respectively turned out to be the best setting for our best performing model CNN-Conext-Rep. The convolutional filter size and pooling size are set to (3, 3). All models are trained with a NVIDIA Titan X GPU using Adam~\cite{DBLP:journals/corr/KingmaB14}. The initial learning rate is 0.001. The parameters of Adam, $\beta_1$ and $\beta_2$ are 0.9 and 0.999 respectively. The batch size is 128. For the word embedding layer, we trained word embeddings with Gensim\footnote{\url{https://radimrehurek.com/gensim/}} with CBOW model using MSDialog (complete set). The dimension of word embedding is 100. Word vectors are set to trainable.
%----------------------------------------
\subsubsection{Evaluation Results}
\label{subsubsec:evaluation-results}
%-------------------------------------------
We select the two strongest feature based classifiers from Section~\ref{sec:baselines} as feature based baselines in addition to neural baselines. They are random forest with the best F1 score and AdaBoost with the best accuracy. The performance comparison of models is presented in Table~\ref{tab:neural-results}.

\begin{table}[ht]
\centering
\footnotesize
\caption{Results comparison. The significance test can only be performed on accuracy. In a multi-label classification setting, accuracy gives a score for each individual sample, while other metrics evaluate the performance over all samples. $\ddagger$ means statistically significant difference over the best baseline with $p < 10^{-4}$ measured by the Student's paired t-test.}
\label{tab:neural-results}
\vspace{-0.3cm}
\begin{tabular}{@{}llllll@{}}
\toprule
Method Types                                                                        & Methods         & Accuracy & Precision       & Recall          & F1             \\ \midrule
\multirow{2}{*}{\begin{tabular}[c]{@{}l@{}}Feature based \\ Baselines\end{tabular}} & Random Forest   & 0.6268                                                          & \textbf{0.7657} & 0.5903          & \textbf{0.6667} \\
                                                                                    & AdaBoost        & \textbf{0.6399}                                                 & 0.7247          & \textbf{0.6030} & 0.6583          \\ \midrule
\multirow{4}{*}{\begin{tabular}[c]{@{}l@{}}Neural \\ Baselines\end{tabular}}     & BiLSTM           & 0.5515                                                          & 0.6284          & 0.5274          & 0.5735          \\
                                                                                    & CNN             & \textbf{0.6364}                                                 & 0.7152          & \textbf{0.6054} & \textbf{0.6558} \\
                                                                                    & CNN-MFS  & 0.6342                                                          & \textbf{0.7308} & 0.5919          & 0.6541          \\
                                                                                    & Char-CNN        & 0.5419                                                          & 0.6350          & 0.4940          & 0.5557          \\ \midrule
\multirow{4}{*}{\begin{tabular}[c]{@{}l@{}}Neural\\ Classifiers\end{tabular}}       & BiLSTM-Context   & 0.6006                                                          & 0.6951          & 0.5640          & 0.6227          \\
                                                                                    & CNN-Feature     & 0.6509                                                          & 0.7619          & 0.6110          & 0.6781          \\
                                                                                    & CNN-Context     & 0.6555                                                          & 0.7577          & 0.6070          & 0.6740          \\
                                                                                    & CNN-Context-Rep & \textbf{0.6885}$^\ddagger$                                                 & \textbf{0.7883} & \textbf{0.6516} & \textbf{0.7134} \\ \bottomrule
\end{tabular}
\end{table}

The base CNN model without feature engineering achieves similar results with the strongest feature based baselines. The performance of the base CNN model is better than the base BiLSTM model. CNN-MFS takes advantage of different filter sizes and also achieves comparable results. Char-CNN, however, performs poorly in this task. Char-CNN does not use pre-trained word embeddings because it learns features from a character-level which would require much more training data.

BiLSTM performs poorly in this task. Even though BiLSTM-Context has a major improvement over BiLSTM, it has inferior results compared to the base CNN model. Compared to non-factoid question answering or chatting, utterances in information-seeking conversations tend to be longer. BiLSTM(-Context) tries to model a holistic sequence dependency and thus performs poorly on handling these long utterances. %CNN based methods look at a window of words every time and thus achieves better results. %Therefore, we build on CNN models for further experiments.

The best result of the feature based baselines is slightly higher than neural baselines. This can be accounted for by the lack of information in neural models. Even though we assume that most of the content and sentiment features can be learned by neural models, the neural models have no access to most of the structural features. Thus, we incorporate all the features to neural models to produce CNN-Feature model. This model outperforms all baseline classifiers. This confirms that incorporating dialog-level information can be beneficial to predicting user intent. 
%Inspired by the performance gain due to extra features, we use two ways to incorporate context information. They are direct expansion (CNN-Context) and context representation (CNN-Context-Rep). \lycomment{These two models are already introduced in Section 5.1.3. }

Both CNN-Context and CNN-Context-Rep outperform baseline models and CNN-Feature without explicit feature engineering. These results demonstrate the effectiveness of the implicit feature learning of neural architectures. CNN-Context-Rep performs better than CNN-Context. This indicates that incorporating high-level features of context information learned by neural architectures is better than directly capturing the raw context information. In a multi-label classification setting, accuracy produces a score for each individual sample, while precision/recall/F1 evaluate the performance over all samples. Thus, accuracy is the only metric that is suitable for significance tests. Our best model, CNN-Context-Rep achieves statistically significant improvement over the best baseline with $p < 10^{-4}$ measured by the Student's paired t-test.

%----------------------------------------
\subsection{Generalization on Ubuntu Dialogs}
\label{subsec:udc}
In this section, we would like to evaluate the generalization performances of different methods on other data in addition to MSDialog. We train different models on MSDialog and test them on the Ubuntu Dialog Corpus (UDC). We select the two best performing feature based classifiers (random forest and AdaBoost) and the best neural model (CNN-Context-Rep) to test the generalization performance. Although the number of annotated Ubuntu dialogs is limited, it is sufficient to demonstrate the predicting performance. We split the annotated UDC data into validation and test sets with an equal size. We train the model on MSDialog data only and tune the hyper-parameters on the UDC validation set. The performance on the UDC test set is presented in Table~\ref{tab:udc-results}. 
% According to a survey by~\citet{5288526} on transfer learning, this setting fall in to the category of \textit{Transductive Transfer Learning}, where the tasks are the same but the domains are different. Although both MSDialog and UDC are in the technical support domain, the language style of the datasets are very different.
% \cqcomment{removed by Bruce}

\begin{table}[ht]
\footnotesize
\centering
\caption{Testing performance on UDC of different models trained with MSDialog. The significance test can only be performed on accuracy. $\ddagger$ means statistically significant difference over both strongest feature based baselines with $p < 0.01$ measured by the Student's paired t-test.}
\label{tab:udc-results}
\vspace{-0.3cm}
\begin{tabular}{lllll}
\toprule
Methods         & Accuracy & Precision       & Recall          & F1             \\ \midrule
Random Forest   & 0.4405                                                         & \textbf{0.6781} & 0.4077          & 0.5092          \\
AdaBoost        & 0.4430                                                         & 0.5913          & 0.4187          & 0.4902          \\
CNN-Context-Rep & \textbf{0.4708}$^\ddagger$                                                & 0.5647          & \textbf{0.5129} & \textbf{0.5375} \\ \bottomrule
\end{tabular}
\end{table}

The generalization results on UDC are not as good as MSDialog. Although MSDialog and UDC both consist of multi-turn information-seeking dialogs from the technical support domain, the drastically different language style adds difficulty for model generalization and transferring. In this challenging setting, CNN-Context-Rep still achieves statistically significant improvement over both baselines in terms of accuracy with $p < 0.01$ measured by the Student's paired t-test.

%------------------------------------------

\subsection{Hyper-parameter Sensitivity Analysis}

We further analyze the impact of two hyper-parameters on CNN-Context-Rep: the number of convolutional filters for the given utterance and the max sequence length. The choices for number of convolutional filters are (64, 128, 256, 512, 1024). We tune the max sequence length in (50, 100, 200, \dots, 1000). As presented in Figure~\ref{fig:tune}, the performance gradually increases as the number of filters increases. The best performance is at 1,024 filters. This confirms our expectation that more convolutional filters can extract richer features and thus produce better results. In addition, the performance fluctuates as the max sequence length increases. Performance with larger ($\geqslant800$) max sequence length are better in general.
% \lycomment{But $300$ is also good ? The average utterance length is only $60$. Such large sequence length is a bit strange.}

\subsection{Case Study}
%\cqcomment{``and error analysis'' removed by Bruce}

Table 10 gives examples of the predictions that different systems fail to make. In the first utterance, the agent asks for the user's iOS version before providing a potential answer, which is a very common pattern of agents' responses. Our CNN-Context-Rep is able to identify the \textit{Information Request} in the utterance while AdaBoost cannot. In the second utterance, both models fail to predict the \textit{Negative Feedback}. This might be due to the fact that the feedback is not explicitly expressed. In addition, it could be relevant that the number of feedback utterances in the training data is relatively limited compared to questions and answers, which makes it more difficult to predict positive/negative feedback.
% \lycomment{Is this due to the small percentage of negative feedback/positive feedback related utterances in the training data ?}

\begin{table}[ht]
\footnotesize
\centering
\caption{Two utterances with their ground-truth and predicted user intent labels. Bold font indicates mispredicted content or labels. ``Ours'' refers to CNN-Context-Rep.}
\label{tab:case-study}
\vspace{-0.3cm}
\begin{tabular}{llll}
\toprule
\multicolumn{4}{l}{\begin{tabular}[c]{@{}l@{}}Hello. Welcome to Skype Community! \textbf{Please provide us the iOS version of} \\ \textbf{your iPad}. The required iOS version for iPad is iOS 8 or higher and for the new \\Skype on iOS requires iOS 9 or higher. For more information, click here. Hope \\ this helps. Let me know if you need further assistance. Thank you!\end{tabular}} \\ \hline
\multicolumn{1}{l|}{Ground truth: \textbf{IR}, PA}                                                            & \multicolumn{1}{l|}{Ours: \textbf{IR}, PA}                                                            & \multicolumn{1}{l|}{AdaBoost: PA}                                                            & Actor: agent                                                           \\ \hline\hline
\multicolumn{4}{l}{\begin{tabular}[c]{@{}l@{}}After modified the Windows entry, value of regedit, \textbf{the error also happened}. \\ When I use C++ for creating another new Microsoft::Office::Interop::PowerPoint::\\Application  instance, the COMException is throwed.\end{tabular}}                                           \\ \hline
\multicolumn{1}{l|}{Ground truth: FD, \textbf{NF}}                                                            & \multicolumn{1}{l|}{Ours: FD}                                                                & \multicolumn{1}{l|}{AdaBoost: FD}                                                                & Actor: user                                                            \\ \hline\bottomrule
\end{tabular}
\end{table}

%----------------------------------------------------------------------------------------
\section{Conclusions} 
\label{sec:conclusion}

% \lycomment{add some sentences on the possible future work here.}
%----------------------------------------------------------------------------------------

% \begin{itemize}
% 	\item Use \textbf{Conclusion} title if only one conclusion
% 	\item If there is no conclusion, use \textbf{Summary}
% 	\item Draw together the topics discussed in the paper.
% 	\item Include a concise statement of the paper's important results and an explanation of their significance.
% 	\item Limitations: shortcomings in the experiments, problems that the theory does not address.
% 	\item Further research or discussion of possible consequences of the results.
% \end{itemize}

In this paper, we studied two approaches to predict user intent in information-seeking conversations. First we use different ML methods with a rich feature set, including the content, structural, and sentiment features. We perform thorough feature importance analysis on both group level and individual level, which shows that structural features contribute most in this prediction task. Given findings from feature analysis, we construct enhanced neural classifiers to incorporate context information for user intent prediction. The enhanced neural model without feature engineering outperforms the baseline models by a large margin. Our findings can provide insights in the important factors of user intent prediction in information-seeking conversations. Future work will consider other methods for user intent prediction. Utilizing the predicted user intent to rank or generate conversation responses in an information-seeking setting is also interesting to explore.
% Future work will consider using the predicted user intent to select or generate more fluent conversation responses in an information-seeking setting.
% \lycomment{Add 1-2 sentences on the future work here.}

\begin{acks}

This work was supported in part by the Center for Intelligent Information Retrieval and in part by NSF IIS-1715095. Any opinions, findings and conclusions or recommendations expressed in this material are those of the authors and do not necessarily reflect those of the sponsor.

\end{acks}

\bibliographystyle{ACM-Reference-Format}
\balance %to have have a more balanced reference list
\bibliography{acmart} 

\appendix
\section{Additional Figure}
\begin{figure}[ht]
	\centering
	\begin{subfigure}[b]{0.24\textwidth}
		\includegraphics[width=\textwidth]{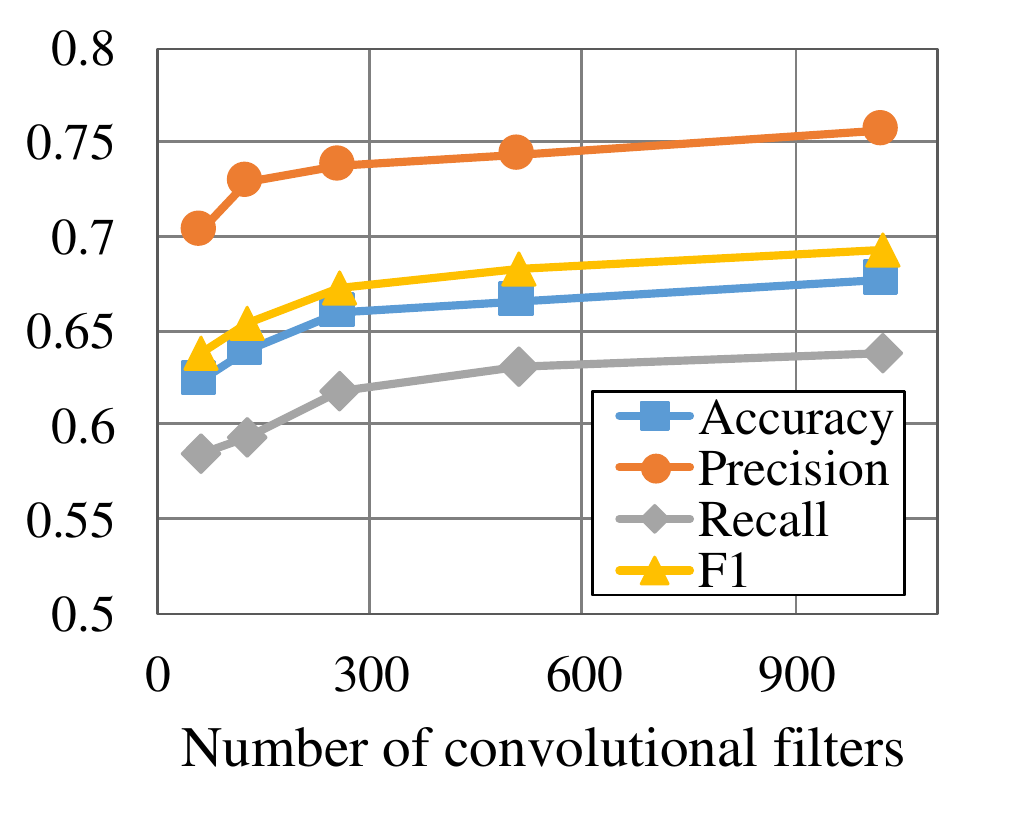}
		\label{fig:nb_filter}
	\end{subfigure}
	\hspace{-0.1in}
	\begin{subfigure}[b]{0.24\textwidth}
        \includegraphics[width=\textwidth]{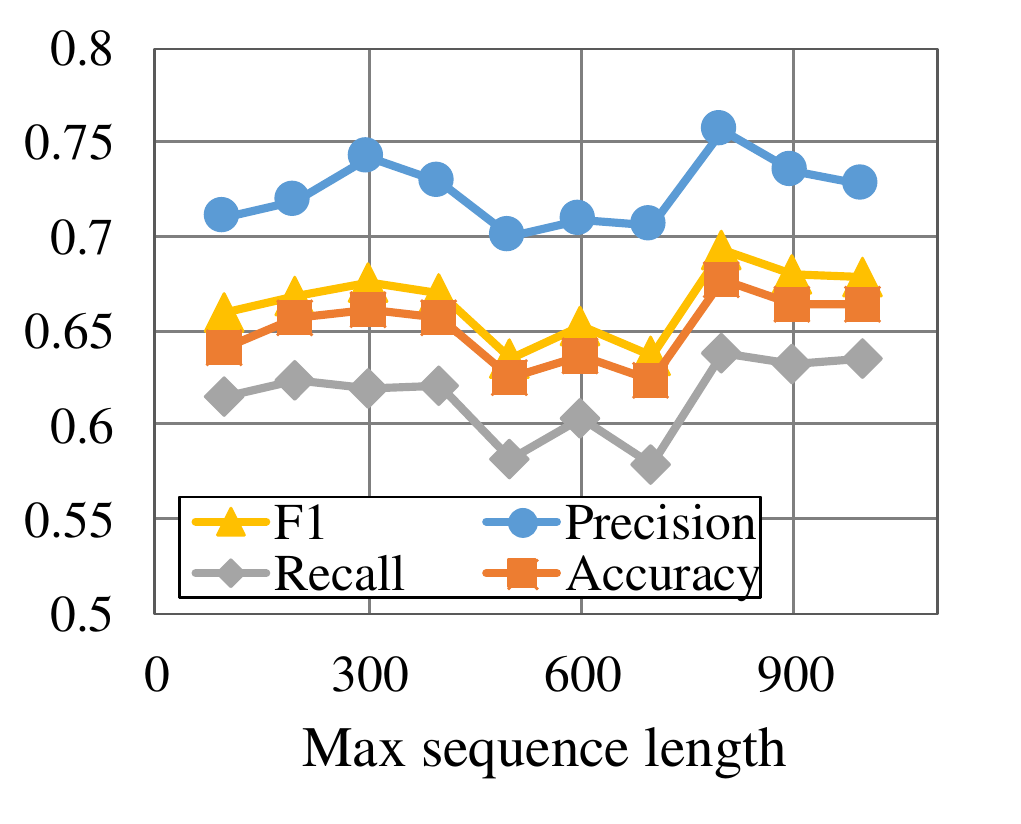}
		\label{fig:max_len}
	\end{subfigure}
	\vspace{-1cm}
	\caption{Performance of CNN-Context-Rep with different number of convolutional filters and max sequence length.}
	\label{fig:tune}
\end{figure}

We include an additional figure to illustrate our findings from above. Figure~\ref{fig:tune} shows the results of the hyper-parameter sensitivity analysis.

\end{document}